# *ParaLog*: Consistent Host-side Logging for Parallel Checkpoints

Steven W. D. Chien
University of Edinburgh
Edinburgh, United Kingdom
steven.chien@ed.ac.uk

Kento Sato
RIKEN R-CCS
Kobe, Japan
kento.sato@riken.jp

Artur Podobas
KTH Royal Institute of Technology
Stockholm, Sweden
podobas@kth.se

Niclas Jansson
KTH Royal Institute of Technology
Stockholm, Sweden
njansson@kth.se

Stefano Markidis
KTH Royal Institute of Technology
Stockholm, Sweden
markidis@kth.se

Micho Honda
University of Edinburgh
Edinburgh, United Kingdom
michio.honda@ed.ac.uk

## Abstract

Output-intensive scientific applications are highly sensitive to low storage throughput. While existing scientific application stacks are optimized for traditional High-Performance Computing (HPC) environments with high remote storage and network bandwidth, these assumptions often fail in modern settings like cloud deployment. This is because the existing scientific application I/O stack fails to leverage the available resources. At the same time, scientific applications exhibit special synchronization and data output requirements that are difficult to satisfy using traditional approaches such as block-level or filesystem-level caching. We introduce ParaLog, a distributed host-side logging approach designed to accelerate scientific applications transparently. ParaLog emphasizes deployability, enabling support for unmodified message passing interface (MPI) applications and implementations while preserving crash consistency semantics. We evaluate ParaLog across traditional HPC, cloud HPC, local clusters, and hybrid environments, demonstrating its capability to reduce end-to-end execution time by 13–26% for popular scientific applications in cloud settings.

## 1 Introduction

Scientific applications are essential for simulating complex physical phenomena, such as weather forecasts, fluid dynamics, and chemical reactions. They traditionally run on high-performance computing (HPC) clusters, but also increasingly on the cloud. Industries, such as automotive [2] and biomedical [51], often run simulations on the cloud if they do not have access to in-house or commercial HPC infrastructures. NIH centers in the US have utilized public cloud resources during the COVID-19 pandemic to run scientific workloads on an unprecedented scale [34]. Despite advances in computing power with heterogeneous systems such as GPUs, managing large data volumes remains challenging. Applications face throughput limitations in both HPC and cloud environments, where storage bandwidth may be scarce. Existing scientific application stacks are designed for HPC clusters in terms of access methods and I/O characteristics, and struggle to adapt to low-bandwidth environments.

Scientific applications are data-intensive and exhibit bursty network output. Climate and weather simulations that study fast-moving processes, such as cyclones, turbulence, and extreme rainfall [86], generate a large amount of data for high-resolution results; combustion or airplane flow simulations, where short-lived phenomena or detailed postprocessing analysis could be missed if the output is infrequent [6, 11, 20, 72, 90]. Output data are stored in a centralized storage system for further analysis by the user. To absorb bursty output, caching is essential to improve an application's end-to-end completion time, but existing approaches fail to achieve essential properties, such as deployability across heterogeneous platforms and crash consistency between the caches across the compute nodes.

ParaLog explores a new design point within distributed logging systems by restricting the I/O subsystem and workload (i.e., write-only workloads with collective synchronization). We believe this is a trade-off worth exploring because it enables a distributed log *consistent across all compute nodes*. It satisfies all requirements that cannot be achieved by traditional methods such as block-level writeback caching [71], or file system journal [35]. The high-level architecture of ParaLog resembles a design pattern found in modern cloud-native databases that combine *shared-nothing* architecture to manage node-local storage devices and *shared-data* architecture to store the primary database in a disaggregated storage [5, 18, 22, 87, 95]. However, focusing on scientific applications, ParaLog deviates from those systems, providing inter-process consistency guarantee based on the MPI-IO calls (§ 6.1 and § 6.5), specialized log management (§ 6.2), and background, yet synchronized checkpointing (§ 6.3).

## 2 Scientific Application Overview

Scientific applications typically consist of two interleaving phases: the compute phase, where distributed processes perform parallel computation; and the output phase, where parallel processes distributed over the compute nodes, together, write the state of the simulation into stable storage over the network. Once launched, they partition and logically distribute the dataset to the processes across multiple compute nodes, and repeat between those phases:

(1) The compute phase by applying mathematical operations to local data, and communicate with others when needed, either in a *point-to-point* (between two processes) or *collective* (all processes participate, e.g., reduce) manner.

(2) After one or more iterations, the processes switch to an output phase, in which the processes write their compute results into the shared files typically hosted in remote storage. The shared file is read by the user(s) for further tasks such as visualization, as input for other applications, or to restart the job after a crash.

After all of the processes complete their remote writes, the processes start another computation phase and repeat.

## 2.1 Storage Architecture

To support concurrent writes, the remote storage typically employs a parallel file system (PFS), such as Lustre [80] and GPFS [68, 79]. Processes can access distinct parts of a file with little contention. Traditional HPC systems deploy PFS as a subsystem. Public cloud operators have enhanced support for scientific applications over the last half-decade in response to the increased demand for them to be run on the cloud. AWS introduced a new managed service, Amazon FSx for Lustre [8], in 2018, because existing storage services, such as Elastic File System, cannot handle parallel write requests issued by scientific applications, and PFSes, including Lustre, are tedious and error-prone for individual users to deploy and operate.

## 2.2 Programming Model

Scientific application processes communicate through the interfaces defined by the Message-Passing Interface (MPI) [31] standard, which offers point-to-point or collective inter-process communication. An *MPI library* implements MPI, and popular ones include Open MPI [84] and MPICH [1]. HPC operators and vendors typically provide MPI toolchains (e.g., Cray Compiling Environment (CCE), Fujitsu MPI, IntelMPI [40]) optimized for specific systems.

While MPI supports the multiple program multiple data (MPMD) model, the common practice is to follow the single program multiple data (SPMD) paradigm. The application is a single binary linked to the MPI library and is launched on multiple nodes through an MPI launcher or job scheduler like Slurm [94]. They then partition the work using information such as the process ID and set up their own parallelization schemes. MPI also provides a set of parallel I/O interfaces called MPI-IO, which coordinates the distributed processes and issues POSIX I/O at each of those to ease parallel file modification by multiple processes.

In the rest of this section, we illustrate the workflow using an example of a climate modeling application with a data layout in Figure 1a. The simulation domain (a global map) is divided into four subarrays, each handled by a process on a separate compute node. During computation, the processes apply mathematical operations to local data and communicate with others when necessary.

During output, these subarrays are written to a shared file, representing the full domain. Since the on-file layout (second row in Figure 1b) differs from the in-memory domain layout (top row), data must be reshuffled into large contiguous blocks to maximize storage throughput and avoid scattered I/O (Figure 1c). MPI-IO facilitates this by allowing each process to define a file view, mapping its local subarray to the global file.

This is done using `MPI_Type_create_subarray`, storing the result in `viewtype`. Each process computes its `global_starts` (offset) and `subarray_sizes` based on its process ID. The processes together open a file in shared storage (e.g., in `/pfs`) using `MPI_File_open`. This is known as a collective operation where all processes must participate. In this example, process 0 writes a 4-byte header with `MPI_File_write_at` (similar to POSIX `pwrite`). Then, all processes collectively set their file view with `MPI_File_set_view` (using `viewtype`, which they created earlier), informing MPI of their access region. Finally, they collectively write their buffer (`my_data`) with `MPI_File_write_all`, letting MPI-IO handle data reshuffling, and issue coordinated I/O requests. The file is then synced and closed collectively.

## 2.3 File Consistency Model

MPI-IO defines the consistency semantics through collective operations such as `MPI_File_sync`. This is conceptually similar to `fsync` but in a distributed manner. Unlike POSIX `fsync`, which is local to a node, `MPI_File_sync` coordinates synchronization across all processes, ensuring the remote file is in a consistent state even with many writers. MPI-IO does not guarantee a remote file is updated unless a `MPI_File_sync` or `MPI_File_close` is performed. Therefore, it is the application's responsibility to order any overlapping write or read-after-write with collective synchronization. We refer to this as the consistency point in the rest of this paper (denoted in red in the figure).

In case of non-storage failure, such as application or node crashing, the file may be corrupted if it has not been synced or closed. For example, if a crash happens before sync in Figure 1c, the state of the file is unclear and thus corrupted. Even after a successful synchronization, if further write operations and a crash happen before the next synchronization or close, the file may again become inconsistent. In this case, while previous data have been safely written to remote storage, the new data remains in an uncertain state. This makes rollback recovery impossible.

# 3 Motivation

Existing cluster architectures such as PFS and communication libraries introduce many optimizations. However, deployment environment – such as on-premise HPC clusters and private cloud – differs significantly, making many optimization assumptions unreliable.

## 3.1 Insufficient Bandwidth

A remote storage must provide the bandwidth required by an application to minimize delay. Due to software overhead and imbalance between components along the I/O path, remote storage can be $1000\times$ slower than compute node's memory bandwidth [85]. For example, the astrophysics code CHIMERA [15] outputs one 160 TB restart file[1] and one 160 TB analytics data per hour in a production environment [49]. To minimize the burst of 320 TB, it requires $320\ TB/s = 2,560,000\ Gbps$ from the PFS. Even PFS on large-scale systems with a large number of backends can fall short of the bandwidth requirement. The periodic bursty output of scientific applications [54, 61, 65] can also be observed in other output-intensive workloads, such as ML training, where the checkpoint frequency is limited by bandwidth [62].

[1] for restarting the simulation from the middle. Checkpoint-restart is the application's responsibility and must not be confused with the log checkpoint in this paper.

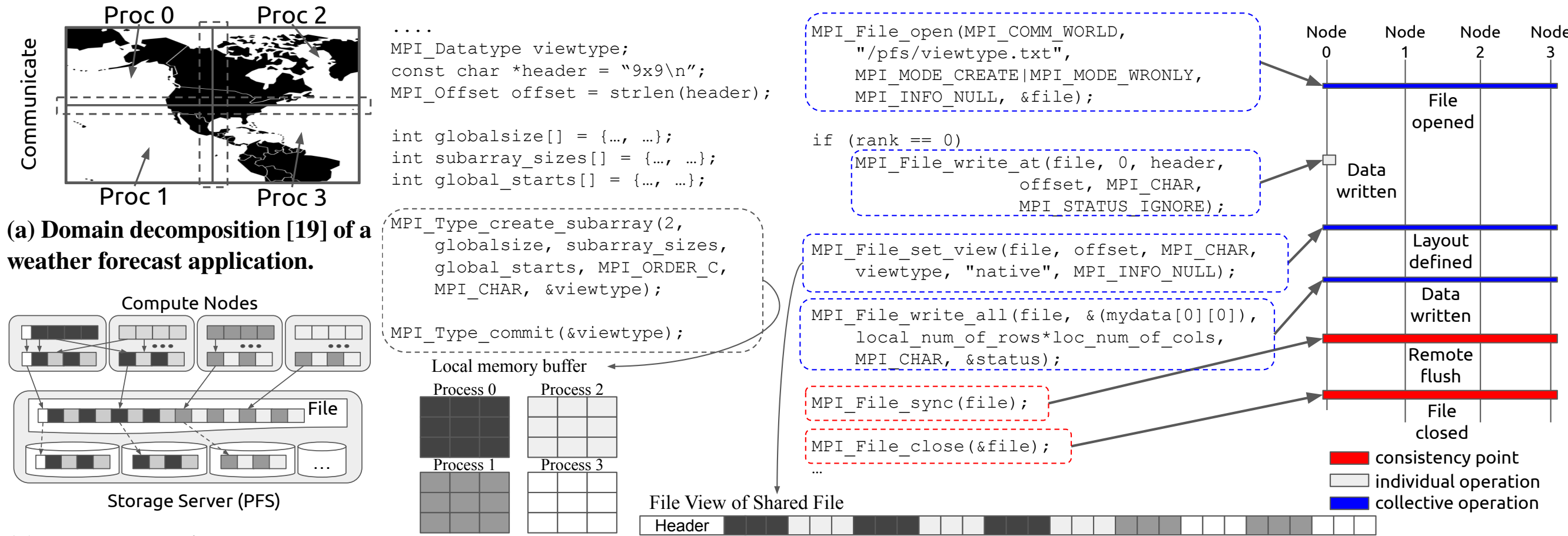

**(a) Domain decomposition [19] of a weather forecast application.**

**(b) Data aggregation and output to a shared file over the network.**

**(c) An example of an application using collective I/O to write subarrays to Figure 1a, together with a header.**

**Figure 1: Scientific application overview.**

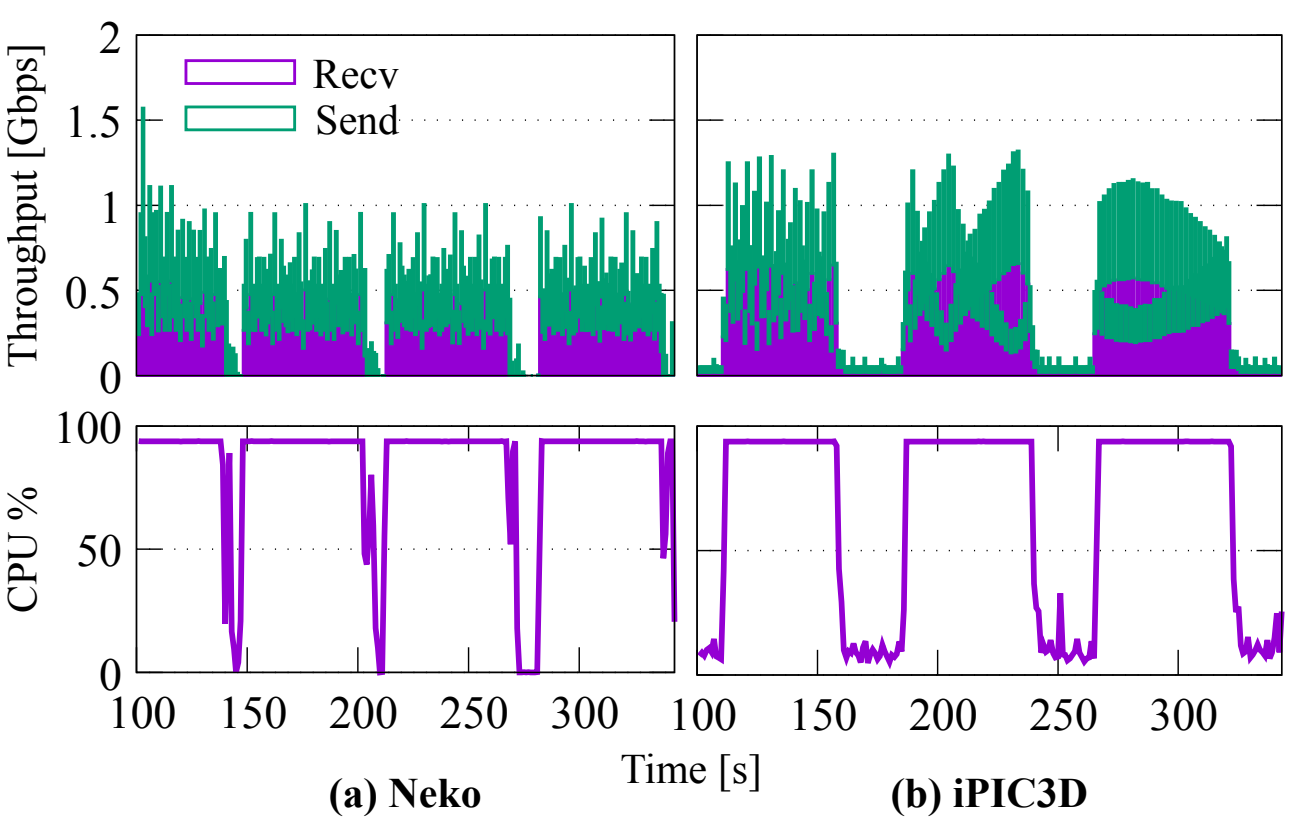

**Figure 2: Inter-node network (top) and CPU (bottom) usage of Neko and iPIC3D with output every minute captured on one node (out of 16). Usage of those drops during transfer to remote PFS (see § 4).**

## 3.2 Platform Mismatch

Deploying a private shared file system on a cloud-based cluster takes considerable system administration effort. For this reason, cloud providers have increased support in recent years and offer *managed PFS* services, such as FSx for Lustre [8] by AWS, and Azure Managed Lustre [10] by Microsoft. However, these systems scale storage and bandwidth by capacity, costing \$12–77.37 per TB per month (FSx for Lustre). To achieve high throughput, the application must over-provision storage space. In our case study (§ 4), eliminating one of our application's output bottlenecks requires nearly 9× more PFS bandwidth, resulting in significantly higher cost and unused storage space. Furthermore, managed PFS is privately provided – unlike HPC PFS, which is time-shared by many users, rendering optimization such as I/O forwarding [3, 44] ineffective.

## 3.3 Operational Mismatch

Managed PFS and cloud instances are private without other users and billed by the hour, leading to wastage if the bandwidth is not constantly used by the application. Techniques such as asynchronous I/O (e.g., `libaio`) can sustain usage, but suffer from overheads [24, 32, 46], constraints such as no buffered I/O, and increased memory usage.

Most importantly, asynchronous I/O does not provide persistence because it uses memory as a buffer. Therefore, it is unsuitable for critical tasks such as checkpointing an application's state. While cloud spot instances offer steep discounts (e.g., AWS spot instances, up to 70% cheaper than on-demand), they risk being recalled with short notice (e.g., 30 seconds), making asynchronous I/O risky. Additionally, the applications' shared-file output pattern makes it difficult to adopt cloud-native alternatives like object storage.

## 4 Case Study

We now present a case study of scientific applications that simultaneously suffer from the above issues (§ 3). We set up a private cloud cluster on AWS using the AWS ParallelCluster tool [7], with 16 compute nodes using `r5.4xlarge` EC2 spot instances. Each is equipped with 16-core Xeon Platinum 8000 vCPU, 128 GB of RAM, and up to 10 Gbps network. We use FSx Lustre as the cluster file system, allocating 6 TiB (minimum required) with HDD storage, giving up to 1.92 Gbps network bandwidth. Our setup costs approximately \$398.44 per month.

We test two applications: the first is Neko [42, 43, 45, 70, 78], a highly-scalable spectral-element code; and the second is iPIC3D [26, 39, 57, 58, 60], a space weather application widely used to study space plasma and magnetic reconnection events. Both applications alternate between blocking compute and output phases, outputting data approximately every minute. During an output phase, the parallel processes coordinate to write their data into a shared-file. We sample the CPU and network usage on one of the compute nodes, as shown in Figure 2. iPIC3D generates dozens of small output files (30–90 MiB) per output, whereas Neko writes one single large file (2 GiB) per output.

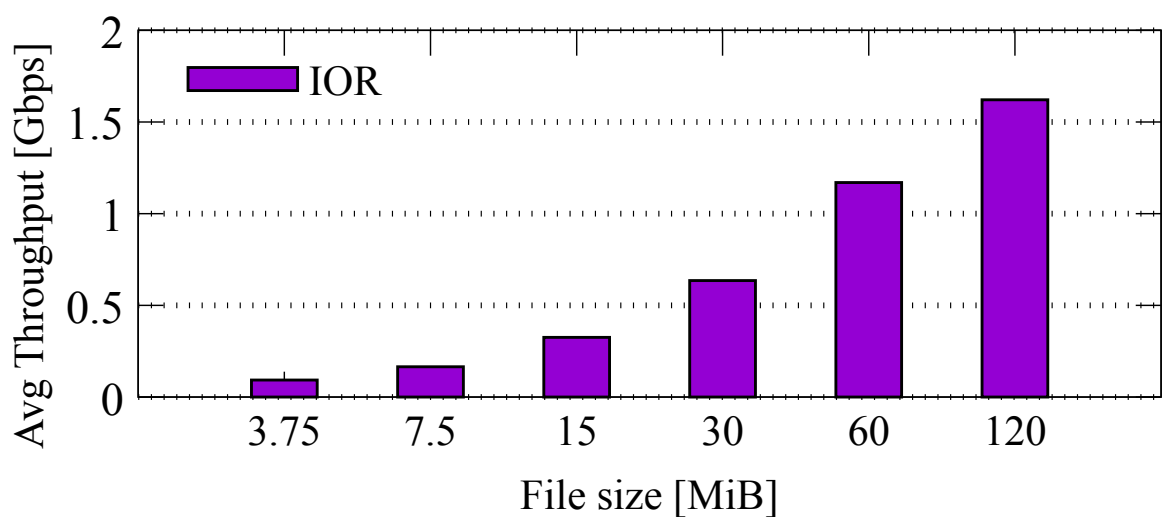

**Figure 3: Throughput of AWS FSx for Lustre (§ 4).**

*Insufficient bandwidth.* While both applications exhibit high CPU usage when computing, they become periodically idle. This is because our setup does not provide sufficient bandwidth. For example, Neko requires $2\ GiB/s = 17.18\ Gbps$, but our managed PFS only gives 1.92 Gbps, leading to a slowdown. This corresponds to a period of low network usage where Neko achieves 1.82 Gbps during output (0.11 Gbps per node, Figure 2:a). iPIC3D only reaches 0.37 Gbps (Figure 2:b), because it writes small files and cannot effectively exploit the PFS' bandwidth. We validate this in Figure 3 by writing files of different sizes using the IOR benchmark tool [81].

*Platform and operational mismatches.* Eliminating Neko's output bottleneck requires nearly $9\times$ more PFS bandwidth, resulting in significantly higher cost and unused storage space. Since the storage bandwidth is not used during computation, it leads to wastage. Without explicit modification to the application, it is impossible to perform asynchronous I/O.

## 5 Design Space

Our example with scientific applications (§ 4) highlights the challenges in deploying applications in a heterogeneous cluster environment. We argue that the fundamental problem is not bandwidth scarcity, but poor utilization. A more practical approach is to transparently buffer data in local persistent storage before spreading remote transfer to the next compute phases [75]. Its impact on the network during compute phases would be minimal, as the bottleneck is on the remote storage. As Figure 2 shows, there is little network usage during output as opposed to the compute phase.

### 5.1 Hierarchical Local Storage

Traditional HPC systems provide local storage via burst buffers, either as node-local SSDs (e.g., Summit [63], Frontier [64], MareNostrum 5 [16]) or near-node storage over dedicated networks (e.g., Cray DataWarp [37], LLIO [28, 76], Rabbit on El Capitan [56]). Similarly, clouds offer virtual block storage (e.g., AWS EBS) with dedicated bandwidth.

Using EBS is cost-effective because the bandwidth of an EBS is independent of its size. For example, AWS's `gp3` EBS provides a base bandwidth of 125 MB/s regardless of volume size ($0.0836/GB-month), whereas our managed FSx Lustre gives 125 MB/s per allocated TiB ($0.151/GB-month). Furthermore, cloud instances already require local block storage to store the operating system – with space not used by the applications. This means that the total bandwidth can scale by the number of compute nodes used, rather than being tied to allocated storage space.

To facilitate applications that write shared files[2], some burst buffers provide a job-local shared-file system mode (e.g, LLIO, Datawarp); while tools like UnifyFS [13], GekkoFS [89], and BurstFS [92], aggregate node-local storage into temporary shared file systems.

Burst buffers are designed for temporary data (e.g., keeping temporary results in MapReduce [23]). They are terminated after job completion and do not persist data to remote storage. As a result, Bez et al. report that 85–98% of jobs in production HPC clusters still rely on PFS as of 2020 [12], citing limited tooling and data staging complexity. Despite this, over 90% of production files are read- or write-only with rare read-after-write [91], indicating the potential for write-only workload optimization.

### 5.2 Caching

Node-local storage (e.g., provided by burst buffers or EBS) can instead be used as a cache to support data staging. However, there are significant deployment and usability challenges. Transparent caching, such as the block-level write-back cache [47, 71], is not suitable for shard-file I/O as it is complex to reconstruct file-level semantics remotely. File system cache approaches such as Arion [35] and LPCC [69] require intrusive kernel and metadata server changes, making them unsuitable for cloud-managed PFS. Furthermore, they struggle with metadata consistency in shared-file I/O. SymphonyFS [63] is a writeback cache which is exposed as a FUSE file system [88]. It starts transferring data from the local cache to the storage server before the application calls sync to reduce the data transfer that blocks the caller at the eventual `fsync` command. Although this approach would work when high remote storage bandwidth is available, as SymphonyFS was designed for HPC platforms, it is inefficient for a cloud environment where the remote storage bandwidth is very slow or expensive. We experimentally validate this in § 8.5.

Caching via I/O forwarding nodes [3, 44] of a PFS requires tight cluster integration, which is unsuitable for small-scale or managed deployments. While it is possible to build them with extra compute instances, this adds cost and underutilizes local resources. Furthermore, I/O forwarding-based optimizations assume a time-shared PFS common HPC with many users, but not in a private cloud environment.

### 5.3 Heterogeneous Storage

Cloud-based PFS lacks elasticity (e.g., the size of FSx Lustre cannot be reduced) and is expensive to deploy. For example, FSx Lustre costs $76.25/month for 6TB HDD, $92.84/month for 1.2TB SSD, and $1,742.44/month for 1.2TB of AWS File Cache. In contrast, cloud-native object stores like AWS S3 are far cheaper at $0.023/GB (1.2TB costs $27.60/month), highlighting the platform mismatch. S3 is reliable, cost-efficient, and widely adopted – even in traditional HPC centers, providing 25% of storage for the Lumi system [21], and 11 out of 54 data repositories in CERN [59]. If applications can use S3, they can completely bypass a slow NFS [66], a tedious and error-prone self-hosted PFS, or an expensive managed PFS.

While S3 offers accessibility [48], its performance and semantics differ from POSIX. For example, objects are immutable and atomic, with no support for ranged writes. Therefore, applications that expect

[2] N-1, i.e., $N$ processes writing one file, as opposed to N-N with file-per-process

a file system have to be significantly re-engineered, from the communication library to data layout (e.g., to use AWS SDK [9]). Tools like s3mount [52] and s3fs [74] expose S3 as a file system but lack multi-writer, shared-file support and cross-node coordination, which causes issues in direct usage. For example, PyTorch encountered non-reproducible errors due to unmet metadata consistency guarantees [73]. Tool such as libCOS [4] enables S3 usage with MPI-IO, but it outputs data in custom-formatted chunks that are linked by a YAML file – not respecting the requested data layout. This makes data dissemination – a strength of using S3 storage – impossible.

## 5.4 Data Safety

The ability to recover unflushed data is crucial for cloud-based applications so that they can exploit low-cost spot instances. Fortunately, cloud block storage, like EBS, is reliable and detached from instances, with failure rates as low as 0.1–0.2% [14].

However, burst buffer file systems like UnifyFS lack crash recovery [55]. File system cache, such as SymphonyFS [63], ensures data safety by blocking at `fsync` until the remote flush is complete. This implies SymphonyFS cannot provide overlapping and crash recovery at the same time. In fact, its benchmark in Ref. [63] does not issue `fsync` during iterations except the last, thus giving no consistency guarantee (§ 8.5).

## 5.5 Deployability

Visibility to application-level synchronization is important as it gives consistency points. Therefore, one approach is to directly implement caching in a communication library – MPI – for the case of scientific applications. Explicit modification of an MPI library [83] is transparent to the application but poses significant deployment challenges due to the variety of MPI implementations, including closed-source and operator optimized ones (e.g., Intel MPI, Fujitsu MPI, Cray Compiler Environment). Deploying ParaLog as a separate MPI library results in users losing access to vendor-optimized libraries, creating a significant barrier.

# 6 ParaLog Design

In each compute node, ParaLog redirects write commands intended for a shared file backed by the remote file system (§ 2.1), to the node-local storage. Each node uses its own local storage, enabling interference-free, exclusive write bandwidth. Those writes are persisted and versioned locally upon an *inter-node* synchronization operation (e.g., `MPI_File_sync`) from the application. The ParaLog data structure is snapshot aware, meaning it supports applications using any synchronization methods, not just MPI. After synchronization points, a snapshot of the remote shared-file can be reconstructed using node-local data on every node. This is an important design decision, as it enables the next compute cycles immediately without completing data transfer, and still supports crash recovery.

## 6.1 Crash Consistency Model

ParaLog preserves the crash consistency expected by ordinary parallel applications (§ 2.3), assuming reliable local storage (§ 5.4) and no storage failure. Remote storage failure follows the same model. To define an inter-process consistency point, ParaLog relies on an I/O library that synchronously issues I/O requests across all the processes, such as MPI-IO.

If a failure happens after a consistency point, the background checkpoint process may be interrupted. However, the checkpoint can be rerun after all the instances are restarted, as all nodes preserve the previous snapshot. This resembles a log redo operation, where logs are replayed after recovering from a crash. In fact, ParaLog provides even stronger crash consistency than direct PFS usage because it only checkpoints a versioned snapshot after a consistency point. Any post-crash writes remain in the ParaLog log as an incomplete record, leaving the remote file uncorrupted.

## 6.2 Local Data Management

*Sequential segment files.* ParaLog keeps data that constitutes a contiguous chunk within the logical remote file in a file backed by a local SSD, which we call a *segment file*. A segment file encodes immutable information, including the remote-file offset and version number (which we call epoch) that identifies the current cycle, in its file name. This is essential, as we will show later, it preserves consistent data snapshots when the application synchronizes or closes the file. Furthermore, fine-grained logging enables flexible remote storage, even non-POSIX ones to be used. ParaLog translates a POSIX I/O syscall (e.g., `write`) based on its file descriptor and offset. When ParaLog captures a `write`, it first checks whether it is contiguous from one of the existing segment files. If it is, that segment file is extended; otherwise, a new segment file is created. Therefore, a remote-file offset is only covered by one file.

*In-memory segment table.* ParaLog maintains an in-memory table that records the names, offsets, and lengths of segment files it created for the same eventual remote file. This table is organized into associative containers (C++ map) sorted by the offset so that ParaLog can quickly search for an existing entry when a write command arrives. ParaLog also tracks the *current* offset in the eventual remote file, because the process can overwrite a part of the existing or current segment file after *seeking* back the offset. Our current implementation only keeps one segment file active (i.e., open `fd`). When a new segment file is created or one of the existing inactive segment files is extended, ParaLog closes the current one.

*On-disk manifest file.* Upon a sync command (e.g., `fsync`), ParaLog closes any active segment file, creates a *manifest* file to store the in-memory segment file table, and persists the manifest file itself. At this point, it is ensured that all the segment files and their metadata (offsets and lengths in the eventual shared file) are durable and consistent. Because this is coordinated by a higher level synchronization (e.g. `MPI_File_sync`), the file state is consistent across all nodes.

*Write reconciliation.* While unusual, overlapping writing can occur, e.g., when an application wants to adjust a header value [91], commonly with I/O libraries such as HDF5 [12, 27, 91]. The MPI-IO standard stipulates linear consistency within a single process. ParaLog's data structure supports this by scanning the in-memory segment table and reusing existing segments. Overlapping segments can be detected (i.e., overwriting the next segment's starting offset) by checking their offset and sizes. The overlapped head of the next segment can be eliminated by an in-place and forward `memmove` of the data, followed by a `ftruncate`. The segment file name is renamed to reflect its starting offset, and the in-memory table needs to be updated.

## 6.3 Checkpoint

Each node runs a checkpoint server in the background and monitors the creation of manifest files, which can be detected by Linux `inotify` or BSD `kqueue`, to transfer the local data to the remote file. Upon signal activation, they read the manifest files, the segments, and write data remotely using regular storage APIs. For example, the checkpoint servers can use MPI-IO to reconstruct the remote file, potentially triggering another round of aggregation for even more contiguous chunks. Since on-disk logs are versioned using the epoch, consistency points do not overwrite each other. Local data is only erased when a remote file has been successfully written.

Some remote storage backends, such as S3, lack a POSIX interface. The current ParaLog implementation directly uses AWS S3 SDK, decoupling data interposition and remote checkpoint. The checkpoint server uses S3 multipart upload for parallel upload, but S3 does not use byte offset semantics and requires the parts to be perfectly aligned. This is problematic if segments are not contiguous (i.e., having holes), or too small (i.e., S3 requires >5 MiB per part).

ParaLog remedy this by writing segment data with MPI-IO through ParaLog again to trigger aggregation for larger and perfectly contiguous chunks. The leader server initializes an upload request and distributes the ID to other servers. Each server uploads its chunks and sends confirmation data (e.g., hash) to the leader. Finally, the leader issues a complete upload request to persist the S3 object. If aggregation fails to produce contiguous segments, all processes send their data to the leader (process zero), which performs a single upload.

In our current implementation, the checkpoint process reads the data from the local disk and writes it back to the network socket, so additional memory usage and copies happen. If this is a concern, they could be eliminated easily using, e.g., `sendfile(2)` or `copy_file_range(2)`, instead of MPI-IO.

## 6.4 Crash Recovery

If a checkpoint is interrupted, it can simply be restarted like a log redo, as the logs are not removed until a successful checkpoint. Since the workflow is equivalent to a remote write using regular MPI-IO or other storage APIs, the performance expectation is equivalent.

## 6.5 MPI-IO Augmentation

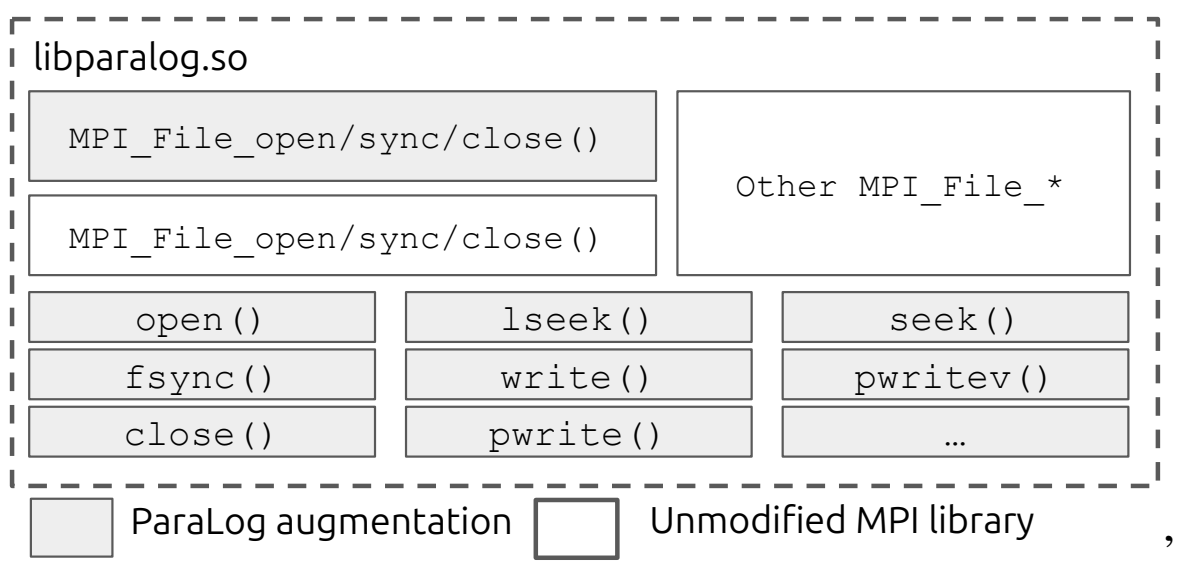

**Figure 4: Design of ParaLog runtime library that augments MPI-IO behavior through preloading selected functions (§ 6.5).**

ParaLog supports unmodified scientific applications that use the MPI-IO interfaces for inter-process communication and storage I/O (§ 2.2). Because of this, it extends generality to also support I/O libraries that are built on top of MPI-IO, such as HDF5 [27] and NetCDF [53].

Since those applications do not directly call POSIX syscalls, such as `open` and `write`, one would build a new fully-fledged MPI-IO library that performs the data management described in § 6.2 behind the standard MPI-IO interfaces. However, this approach is not viable, because it requires implementation of the full MPI-IO abstraction layer, including read functions, which is out of the scope of ParaLog. More importantly, many scientific application deployments opt for platform-optimized, proprietary MPI-IO libraries whose source code is unavailable to the user or operator. If we sacrificed the choice of the MPI library, the use of ParaLog would be a huge burden for users.

ParaLog therefore *augments* the existing, unmodified MPI-IO library as in Figure 4. It intercepts the MPI-IO calls that define inter-process consistency points across the compute nodes, which are `MPI_File_open`, `MPI_File_sync` and `MPI_File_close`, by *preloading*. ParaLog also intercepts POSIX syscalls issued by the MPI-IO library to manipulate the segment files and segment tables discussed in § 6.2. When `MPI_File_open` issues the POSIX `open`, which is preloaded, ParaLog returns a *placeholder* file descriptor to the MPI-IO library.

Since the MPI-IO library uses this descriptor for all the further syscalls on this file, ParaLog can identify the syscalls originated by MPI-IO. To ensure that the placeholder descriptor number is unique, ParaLog opens a temporary file to obtain and occupy this number and maintains it in a hash table.

Importantly, ParaLog only intercepts `MPI_File_open` and the subsequent `open` call that matches a path prefix provided by the user through an environment variable, and opened in write-only mode. ParaLog would not interfere with any other file operations, or those issued by itself.

## 6.6 Implementation and Deployment

The ParaLog preloaded library, checkpoint server, and recovery tool consist of fewer than 1.5k LoC of C/C++ code. We validated that ParaLog can be used with the following *unmodified* popular open source MPI libraries: Open MPI, MPICH, ParaStation MPI; and closed source ones: Cray MPI, Fujitsu MPI, Intel MPI. In addition to preloading the ParaLog library and setting relevant paths, the checkpoint server must start in the background with one server per node, specifying a remote checkpoint location. The code is available on a repository[3] which can be installed using Spack [29]. Since ParaLog uses Linux's preloading, it does not modify or reverse engineer any closed-source binaries. Using the ParaLog library itself does not create a license issue, as linkage to the application is not required.

# 7 ParaLog in Action

ParaLog is activated by preloading (`LD_PRELOAD`) an augmented library and pointing to the file path to the local SSD while keeping the remote file name. Those are set in environment variables. In this section, we detail how ParaLog's data management works, referring to Figure 5.

[3]https://github.com/uoenoplab/ParaLog.git

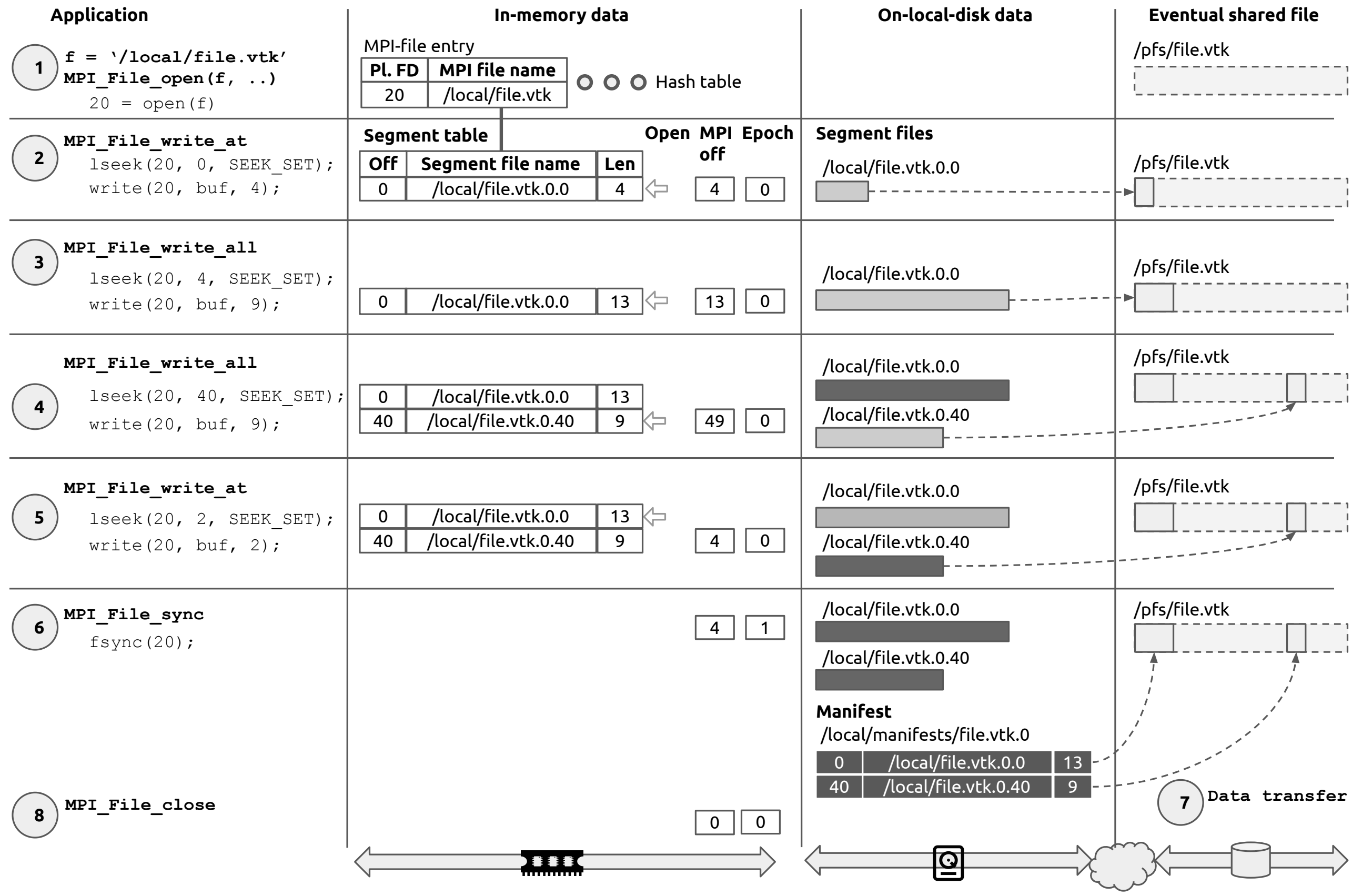

**Figure 5: Logging operations triggered by application calling MPI-IO functions, and their underlying POSIX I/O functions. ParaLog only captures `MPI_File_open`, `MPI_File_sync`, and `MPI_File_close` for consistency points, and POSIX I/O functions for data and file pointer operations (§ 7).**

① **Open an eventual shared file.** When the application opens (`MPI_-File_Open`) a new eventual shared file (`file.vtk`), but specifying prefix `/local`, the MPI implementation issues POSIX `open` on that path. ParaLog intercepts this open if it is in write-only mode and the prefix matches with the environment variable, returning a placeholder descriptor (`fd`=20). ParaLog registers it to the hash table that maintains the MPI-file entries and the segment table is initialized (§ 6.2).

② **First write.** Consider the application calls `MPI_File_write_at` to write data at a specific position in the target file (executed by an individual process). `MPI_File_write_at` first `lseek` to 0, then `write` 4 bytes of data. ParaLog intercepts those calls as they are issued on the placeholder `fd` (20). When ParaLog sees `lseek`, it records the specified offset (MPI off in the figure). When ParaLog sees `write`, since there exists no segment table entry, it opens a new segment file, `/local/file.vtk.0.0`, and stores the written data. Note that POSIX calls issued by ParaLog itself are not intercepted. ParaLog then updates the offset (again MPI off in the figure) to 4, as 4 bytes of data have been written at offset 0. ParaLog records the remote offset, name, and length of this segment file in the segment table.

③ **Contiguous write.** The application then calls collective `MPI_-File_write_all`, where every process issues a write at a position specific to each of them. The MPI implementation `lseek` the offset to 4, followed by `write` with 9 bytes of data. Since the seek offset matches the current one that was updated by the previous `write`, ParaLog appends 9 bytes of the data to the existing segment file and updates the length field of the segment table entry and the current offset (MPI off) to 13.

④ **Discontiguous write** In another `MPI_File_write_all`, consider that the process moves the MPI-file offset to 40 and writes 9 bytes of data. This is discontiguous from the last write position, which is 13, and there is no other segment file that this write can extend. ParaLog closes the currently open segment file and creates a new one `/local/file.vtk.0.40`. A new segment table entry is inserted to record this segment file name and offset (i.e., 40), as well as updating the current offset (MPI off) to 40 (which is further updated to 49 after `write`). ParaLog then writes the data to the new segment file and updates the length field of the segment table entry.

⑤ **Overwrite.** The MPI implementation then writes 2 bytes of data at offset 2. Since this position is inside the existing, inactive segment file (`file.vtk.0.0`), ParaLog closes and persists the current open segment file and reopens that one. It then sets the offset to 2, writes 2 bytes of data, and updates the current offset to 4. Note that this `write` does not update the length field of the segment file entry.

⑥ **Sync.** When the application calls `MPI_File_sync`, ParaLog persists the active segment file (`/local/file.vtk.0.0`). It then creates a manifest file, `/local/manifests/file.vtk.0`. This file records the name, offset, and length of the two segment files that have been created over the epoch. The manifest file is then persisted, which is detected by the checkpoint server (§ 6.3). Finally, ParaLog updates the epoch.

⑦ **Data transfer** The checkpoint servers on each node pick up the signal when a manifest file (`/local/metadata/file.vtk.0`) is committed to local storage. They parse the manifest, read segment files into memory, and collectively reconstruct remote data. If PFS is used, the servers use regular MPI-IO to describe layouts and write data. The segment `/local/file.vtk.0.0` is written to byte 0–13, and `/local/file.vtk.0.40` to byte 40–49 in the remote file (`/pfs/file.vtk`). If S3 is used, the servers perform the same operation but use `/local` rather than `/pfs` as the output target. This triggers a further aggregation of segments and output through ParaLog. The output file name is munged with a suffix (e.g., `#FOR_-S3#`) to avoid overwriting the consistency point. The servers send their re-aggregated segment offsets and lengths to a leader. The leader sorts and verifies that all segments are contiguous and distributes part numbers back to them. For example, `/local/file.vtk#FOR_-S3#.0.0` gets assigned part number one. The servers proceed to upload the segments individually and send confirmation information to the leader. Finally, the leader issues a completion request, and the S3 object is persisted. Local segment and manifest files are cleaned up after all the servers have completed their transfer.

⑧ **Close.** When the application closes the shared file, ParaLog deletes all the in-memory data. However, data transfer to the remote storage can still be in progress; the checkpoint server is responsible for deleting the segment files and manifests. This is because all the logs required for reconstruction persist on the nodes' local storage.

## 8 Evaluation

To demonstrate the wide applicability of ParaLog, we use five systems (Table 1) to evaluate different aspects of ParaLog. We reserve one core per node to run the checkpoint server and use the rest for the application[4].

### 8.1 Traditional HPC Systems

How does ParaLog perform in traditional HPC systems? We use two HPC clusters. The first is Vega [41], which is a petascale supercomputer with AMD CPUs and NVMe SSDs. The system is accessible to expert researchers in Europe to run scientific workloads that include fluid dynamics, astrophysics, and ML. The second is NextgenIO [67], which is an early-stage HPC system with Intel CPUs and Optane Persistent Memory storage on each node to support data-intensive applications. We use Neko as the application because it can easily scale to arbitrary numbers of cores without altering decomposition schemes between those clusters. Neko exhibits alternating compute and checkpoint phases, as many other scientific applications (§ 4).

Figure 6 plots the end-to-end job completion time with a strong-scaling (fixed problem size) setting. ParaLog has little effect on two nodes because of little storage or network contention. From 4–16 nodes, ParaLog outperforms the baselines by a larger margin with increasing output frequency. For example, when the application outputs every 20 cycles, we observe 36.6–64.3% improvement in Vega and 16.5–17.5% in NextgenIO; when doing so at every 4 cycles, 58–90% in Vega and 33–44% in NextgenIO, depending on the number of nodes. ParaLog achieves better improvement on Vega than NextgetIO, because it experiences more PFS contention in the former. We conclude that our results show that ParaLog effectively caches write operations of individual nodes and spreads remote storage writes over compute phases in traditional HPC clusters.

### 8.2 FSx for Lustre in Public Cloud

How does ParaLog accelerate scientific applications executed in public clouds? We run Neko and iPIC3D on the **AWS cluster** setup described in § 4, and Figure 7 plots the results. Similarly to the HPC clusters used in the previous subsection, ParaLog reduces the end-to-end job completion time by a larger margin with increasing output frequency in Neko (2.5–40%) and iPIC3D (6.8–38.6%) with Intel MPI. We also observed similar results with Open MPI, improving 4.86–58.14% in Neko and 3.7–68.7% in iPIC3D. ParaLog reduces required throughput by spreading remote PFS writes to avoid bursty bandwidth demand.

Without ParaLog, iPIC3D exhibits a longer idle period between compute cycles (Figure 2) because it generates many smaller files that cannot exploit PFS bandwidth. Therefore, ParaLog alleviates not only network bandwidth shortage (Neko) but also storage bandwidth shortage (iPIC3D).

### 8.3 S3 Backend

Without ParaLog, users are expected to first write data to PFS using regular MPI-IO and then upload it to S3 using a tool like `s3cmd`. To the best of our knowledge, ours is the first work that examines a fully-fledged MPI-IO application that directly writes data to S3 in a production HPC system.

We first run ParaLog with Neko on Lumi, scaling up to 32 nodes. Since Lumi compute nodes are diskless, we use a ramdisk (backed by `ext4`) for logging, thus providing no crash consistency in this experiment. In Figure 9, ParaLog is slower than PFS for infrequent outputs (20 or more cycles per output) but outperforms PFS with more frequent outputs, achieving up to a 28% reduction at four cycles per output.

In applications such as weather forecasts, computation results are distributed to remote users as soon as they are available. To explore Internet dissemination, we run iPIC3D and WRF on **ICSY**, checkpointing data to Spaces Object Store[5]. WRF [82] is a widely used weather forecast code that writes NetCDF [53] files periodically [38]. We use the 12km variant (approximately 500 MiB per output), due to the limited space (which we can afford) of S3. We compare the performance of S3 output cases against using an NFS server which connects to **ICSY** over a 25 Gbps link (§ 8.4).

[4]Whenever Lustre is used, we set a stripe count of -1 to ensure all the storage backends are used. For local storage, we use XFS and ext4 on HPC systems; ext4 on on-premise cluster and for ramdisks.

[5]An S3-compatible cloud storage hosted by Digital Ocean, approximately 112 MiB/s for upload, measured using **warp** https://github.com/minio/warp from one node.

**Table 1: Evaluation setup.**

| System | CPUs and Total Cores per Node | RAM | Node-Local Storage | Baseline Cluster Storage | Network | Toolchain |
|---|---|---|---|---|---|---|
| **Vega** | 2x EPYC 7H12 (128 cores) | 256GB | Micron 7300 1.92TB | Lustre | InfiniBand HDR | Open MPI, GCC |
| **NextgenIO** | 2x Xeon Platinum 8260 (48 cores) | 192GB | NVDIMM AppDirect 3GB | Lustre | Omnipath | Open MPI+GCC |
| **Lumi** | 2x EPYC 7763 (128 cores) | 256GB | Ramdisk (diskless) | Lustre, Lumi-O (S3) | Slingshot-11 | Cray MPICH+CC |
| **AWS cluster** (16 nodes) | Xeon Platinum 8000 vCPU (16 cores) | 128GB | Elastic Block Storage 125MB/s | FSx Lustre | Up to 10Gbps | Intel MPI+ICC, Open MPI+GCC |
| **ICSY** (5 nodes) | 2x Xeon Silver 4314 (32 cores) | 128GB | Samsung PM9A3 NVMe SSD | NFS, Digital Ocean Spaces (S3) | 25Gbps Ethernet | Intel MPI+ICC, Open MPI+GCC |

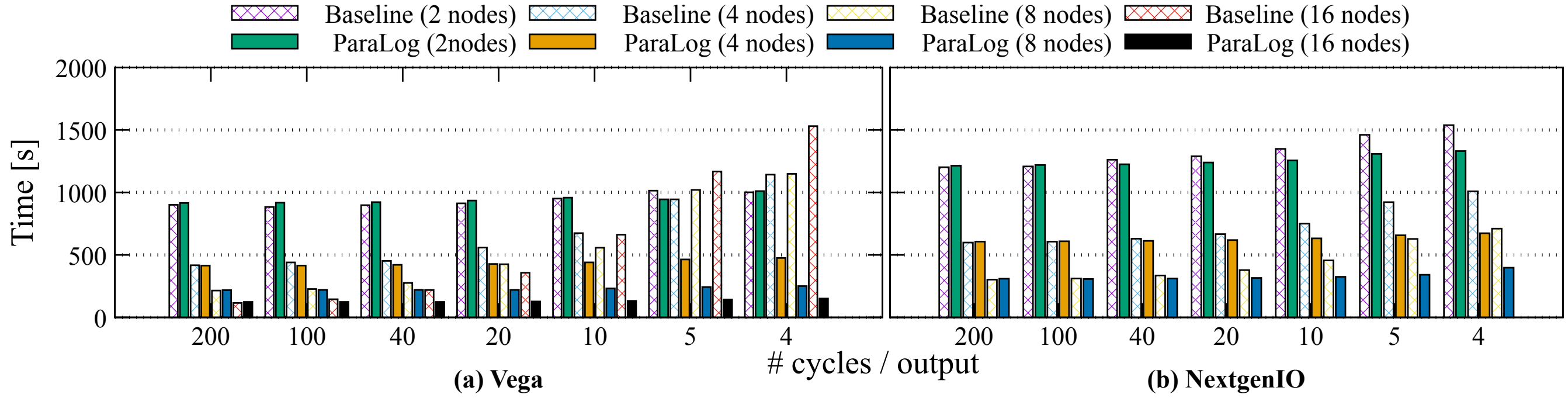

**Figure 6: End-to-end completion time of Neko on HPC systems (§ 8.1) (up to eight nodes for NextgenIO).**

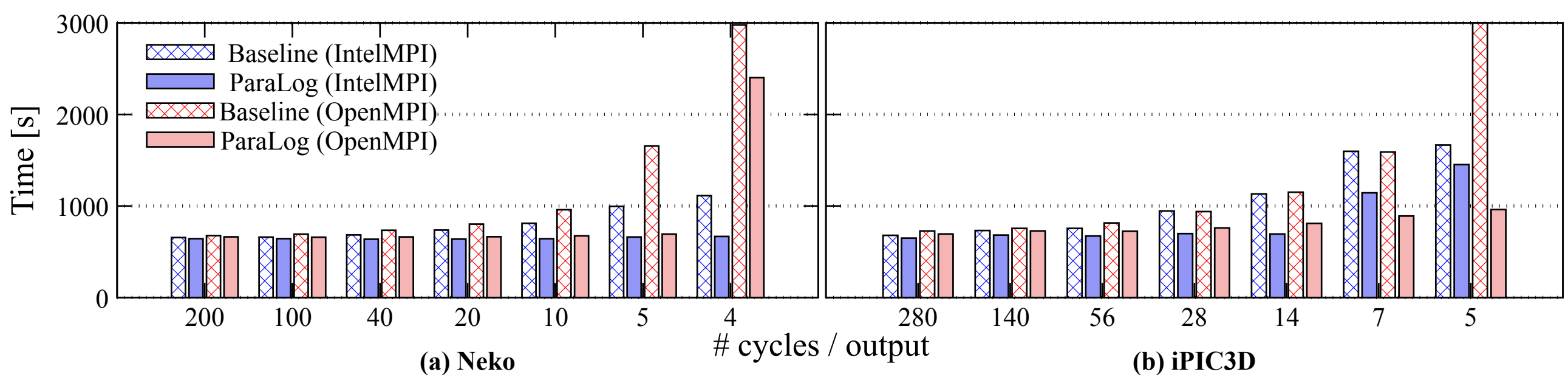

**Figure 7: End-to-end completion time on AWS cluster (§ 8.2).**

Figure 8:b,d (Baseline and ParaLog-S3; we will discuss ParaLog-NFS in the next subsection § 8.4) show the results where output to S3 exhibits comparable performance to outputting directly to NFS, despite each node in **ICSY** is connected to the Internet through a 1 Gbps link. Our results demonstrate that ParaLog efficiently utilizes low network bandwidth, allowing new workflows where scientific applications decouple computation from the storage, even geographically; and enabling users to instantly access output stored in ordinary cloud storage.

## 8.4 NFS Backend

Inspired by previous experiments, some users might prefer to use ordinary NFS, especially in existing clusters that already deploy it. To test this scenario, we use **ICSY** with five compute nodes. Figure 8:a,b plot the results, in addition to ParaLog-S3 which was discussed in § 8.3. At the highest frequency, ParaLog reduces the end-to-end completion time of Neko by 44-49%; and iPIC3D by 27-31%. As with **AWS cluster**, ParaLog gives little or no benefit at very low frequency.

We also run WRF. In addition to the small 12km variant (same as the previous subsection), we run the larger 2.5km variant, which writes approximately 8 GiB per output. Figure 8:c and Figure 8:d show that ParaLog achieves improvement of 11–49% with Intel MPI and 8–54% with Open MPI, depending on the output frequency.

Our results show that ParaLog can accelerate end-to-end job completion time over a wide range of output frequencies (in best cases, up to 58% and 68% for Neko and iPIC3D, respectively). Notably, ParaLog achieves compute-output overlapping irrespective of MPI implementations in both HPC, on-premise clusters, and the cloud, demonstrating its wide applicability.

## 8.5 Comparison to SymphonyFS

We compare ParaLog against SymphonyFS, which is a writeback cache that is the most similar to ParaLog to our knowledge. It supports arbitrary remote POSIX file storage with write-back caching at the node-local block device (§ 5.2). While SymphonyFS must block on `fsync` until remote sync is complete, ParaLog can return immediately once data is persisted locally. This is because ParaLog provides crash recovery from local logs even if remote data is not completely reconstructed. Whereas SymphonyFS has no logging or cross-process synchronization knowledge. Another important implication is that SymphonyFS cannot support non-POSIX semantics such as S3. ParaLog only initiates a remote checkpoint when data becomes immutable at a consistency point, which can effectively resolve semantic differences between POSIX and object storage.

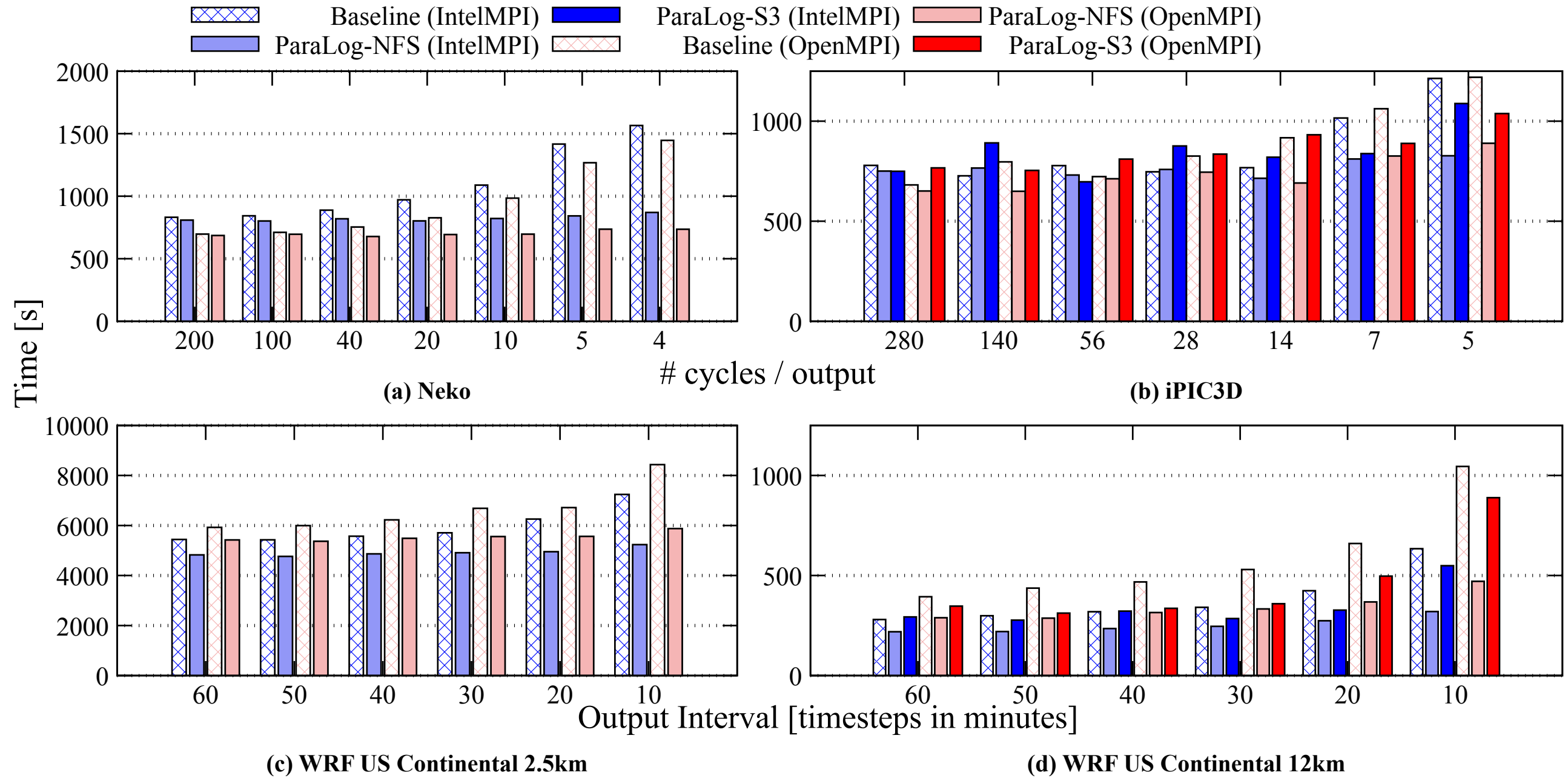

**Figure 8: End-to-end completion time on ICSY (§ 8.3 and § 8.4).**

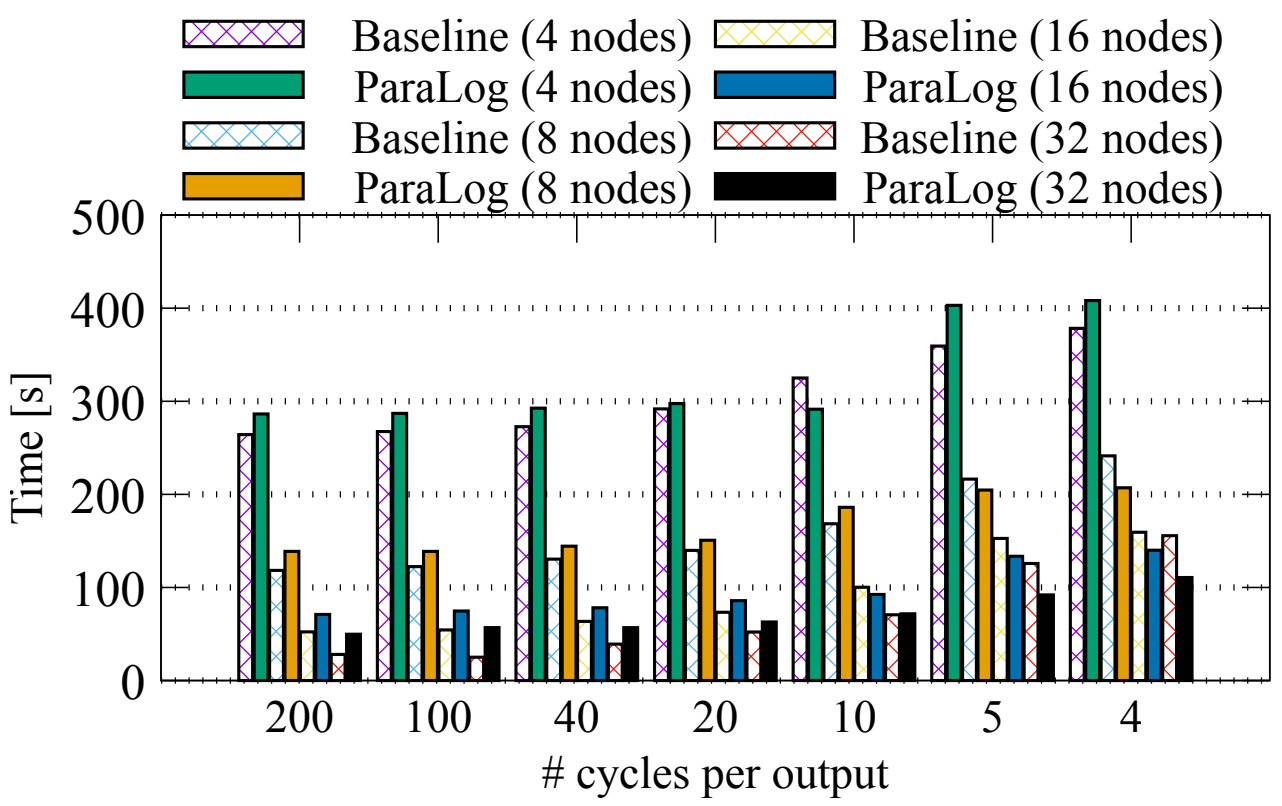

**Figure 9: End-to-end completion of Neko on Lumi uploading to Lumi-O S3 through ParaLog buffering on ramdisk (§ 8.3).**

We use **ICSY**, because SymphonyFS requires the root privilege to use FUSE. Figure 10a plots the results with the IOR benchmark [81] which emulates scientific applications producing different output sizes. ParaLog performs the best except for the case with 20 small outputs over 5 minutes. This is expected; since ParaLog persists data and metadata of every `write`, the frequent output of small files exhibits the worst case. However, overall, ParaLog achieves a similar or better performance than SymphonyFS while guaranteeing crash consistency.

After that, we run a real-world application, Neko (Figure 10b). To the best of our knowledge, SymphonyFS has not been evaluated against real-world workloads in literature [83], and Ref.[63] only performed IOR experiments. We control the network bandwidth with Linux `tc` to emulate throughput limits and eliminate the NFS server disk bottleneck using a ramdisk. As expected, when the remote storage bandwidth is lower, ParaLog outperforms SymphonyFS by a larger margin (up to 23%), demonstrating the advantage of local sync over earlier remote sync. Furthermore, the application cannot enforce synchronization in SymphonyFS without blocking the application, which means consistency is not supported (thus no consistency).

## 8.6 Data Recovery Performance

Since a checkpoint process is equal to a remote write using regular storage APIs (e.g., MPI-IO), re-running a checkpoint gives a similar performance. We measure file recovery performance by populating the output files of a single iteration of Neko and iPIC3D in local storage. We then measure the time taken to transfer (i.e., recover) those files into the remote aggregated file in FSx Lustre. As expected, results in Figure 11 show that throughput is constrained by the remote server, similar to observations in Figure 2.

When an application restarts, it fetches restart files from remote storage. ParaLog would cause extra delay in restarting only if the latest checkpoint has been interrupted. In this case, the checkpoint has to be rerun before the application can restart.

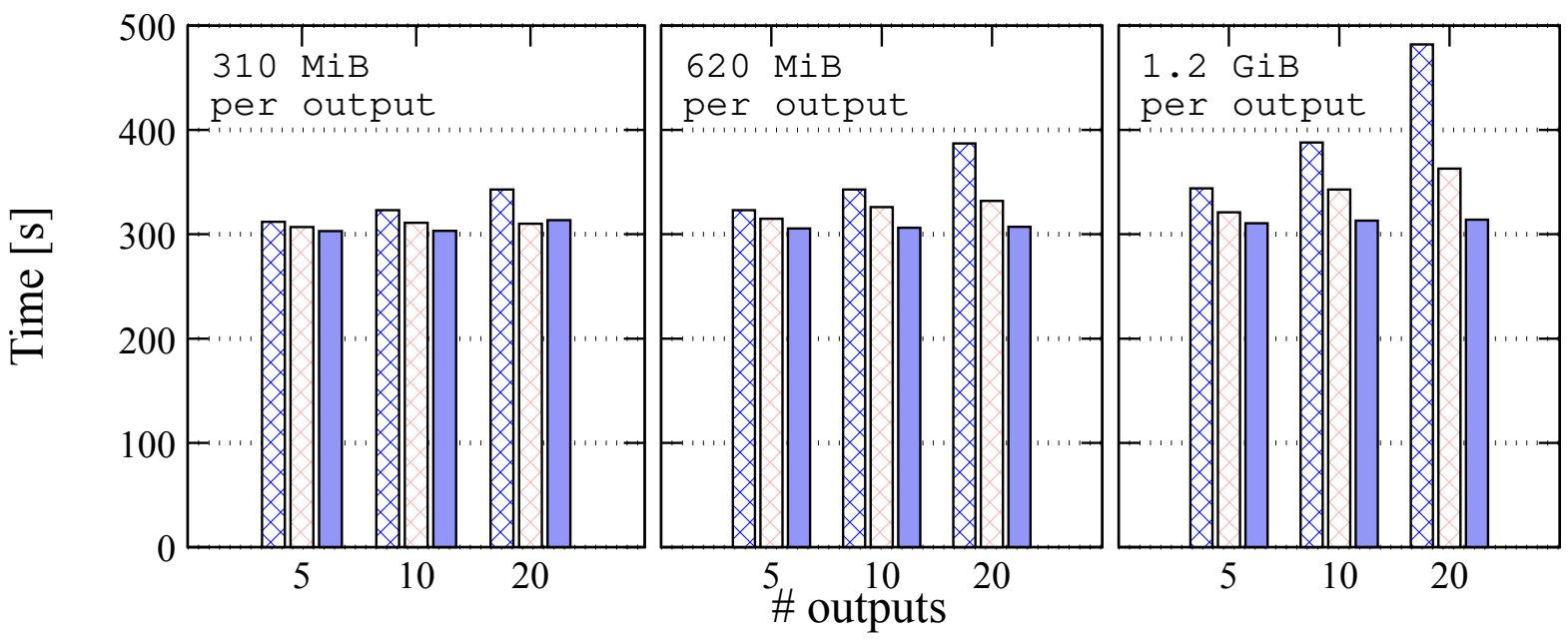

**(a) Acceleration of IOR benchmark using SymphonyFS and ParaLog over NFS. Each experiment "computes" a total of five minutes and outputs between 5–20 times. The NFS is backed by a SATA HDD.**

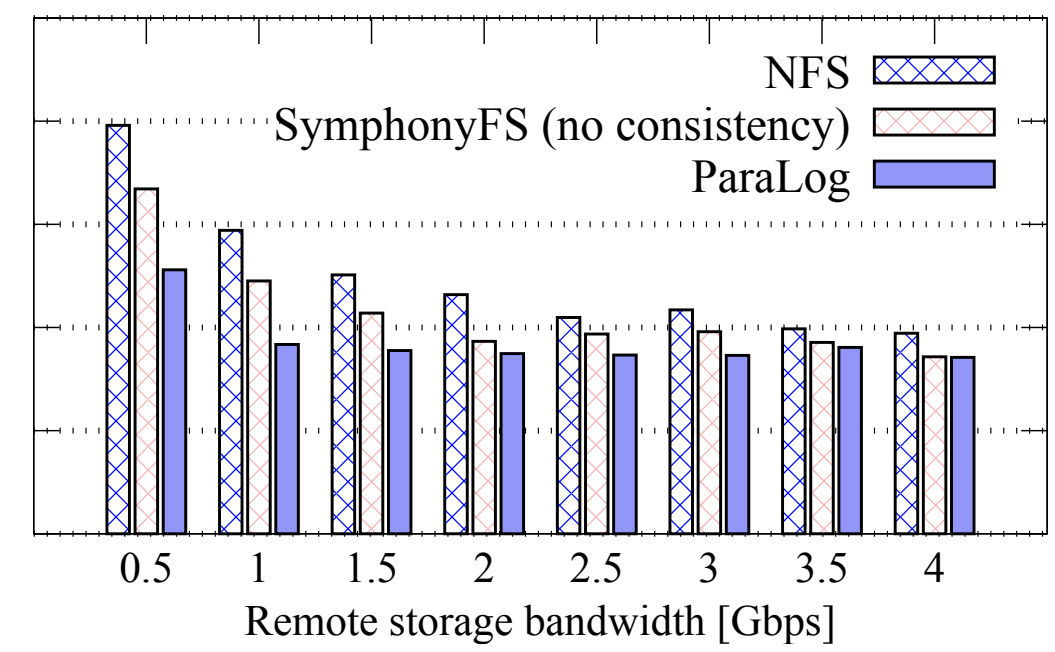

**(b) Neko outputting data under different bandwidth limits.**

**Figure 10: Comparing between ParaLog and SymphonyFS on ICSY (§ 8.5).**

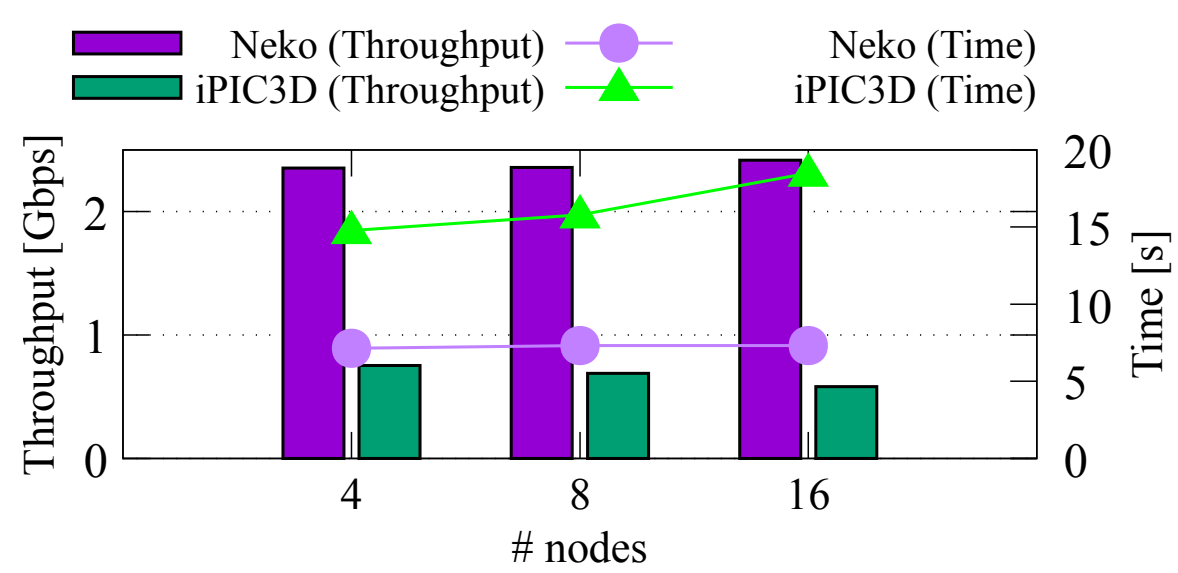

**Figure 11: Recovery performance, where lines are associated with time, and bars are associated with throughput. (§ 8.6).**

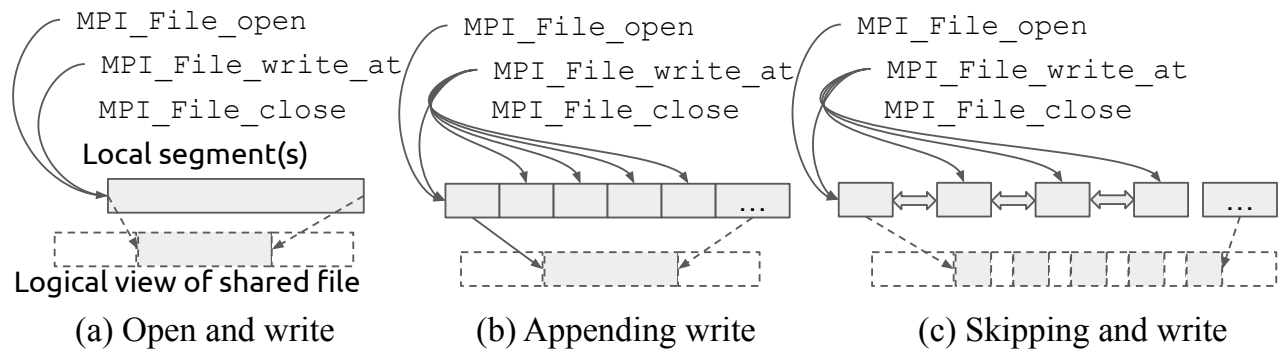

**Figure 12: Tool to stress the ParaLog write cache management (§ 8.7).**

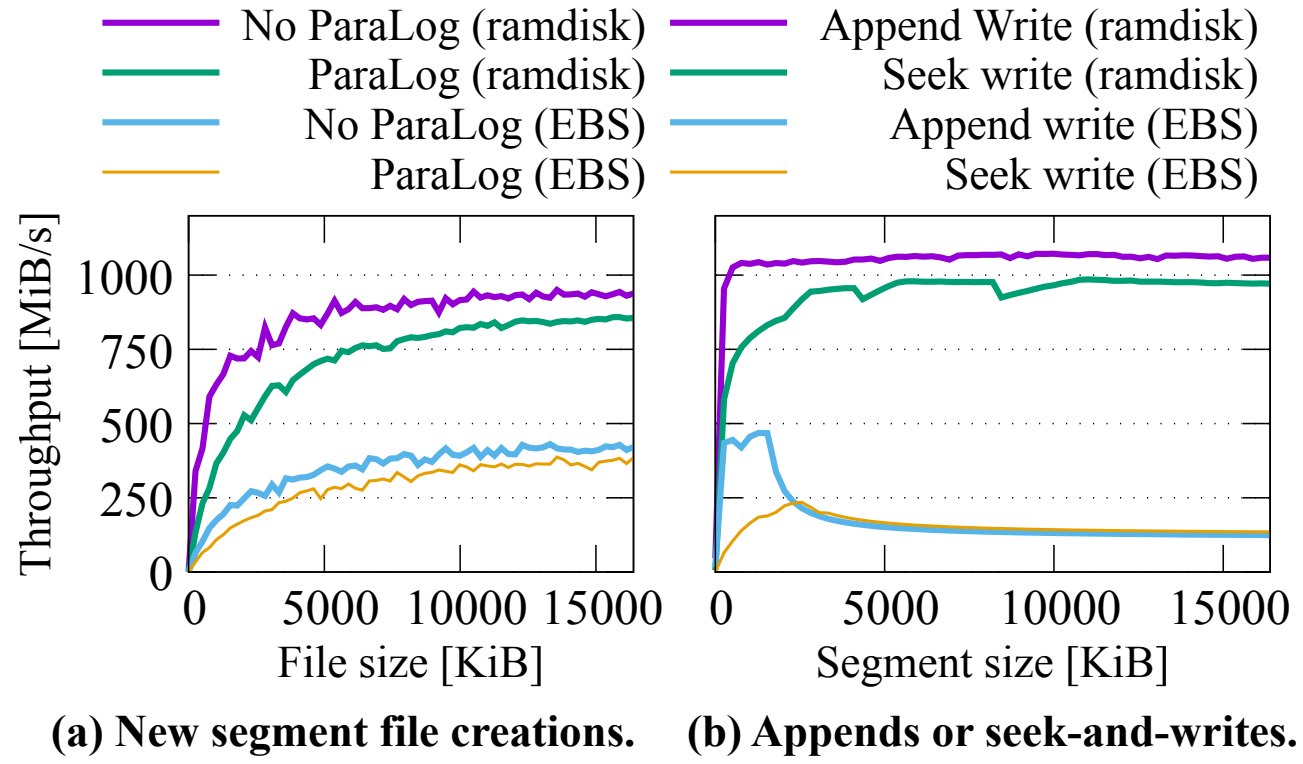

**(a) New segment file creations.** **(b) Appends or seek-and-writes.**

**Figure 13: ParaLog local data management microbenchmarks (§ 8.7).**

## 8.7 Local Data Management

We have so far demonstrated how ParaLog can exploit systems with different performance characteristics. However, would ParaLog still be effective with extremely fast local storage? To answer this question, we measure the performance of our write cache management described in § 6.2 on one process in one node on our **AWS cluster** to quantify its overhead.

*8.7.1 Open and Write.* Since we create a segment file for every write operation unless it can extend the previous one without a gap, the extra costs in creating and closing a file can lead to overhead. We characterize those overheads using a custom program that opens, writes one segment of data (1–16384 KiB), and closes a file (Figure 12:a); this I/O pattern resembles the iPIC3D application. We use an EBS block device and a RAMDisk, emulating an extremely fast block device to highlight the file metadata management overhead of ParaLog.

Figure 13:a plots the results, where the baseline refers to writing without ParaLog. The relative overheads are amortized over the movement of larger data segments. This convergence (between the darker and lighter lines) is even clearer on the ramdisk. We note that the throughput converges to a higher rate than the EBS baseline bandwidth (125 MB/s) as they likely benefit from the burst allowance. We conclude that the per-write overheads in ParaLog are low enough, even when the backing block device is very fast.

*8.7.2 Contiguous and Discontiguous Writes.* We then benchmark the performance of the other write pattern, which extends existing segments (Figure 5:③) or writes in a new offset in the eventual shared file (Figure 5:④). We keep the file but continuously write 100 new segments into it (i.e., writing to where the last write left off).

Figure 13:b plots the results. As expected, writing to a new offset (with a seek) is significantly slower than extending the existing writes, because it needs to flush the current segment file and create a new one. However, the costs of those operations are amortized when the segment size is large enough (e.g., 4096KiB), resulting in a similar throughput to the append writes. We again attribute the initial high rate in the Append-write case for EBS to the benefit of burst credit.

Our results show that ParaLog's I/O interposition introduces minimal overhead in a range of storage devices. Furthermore, our results on EBS show that cloud block storage is fast enough even though they are decoupled from the instances. We note that segment sizes are typically large because MPI-IO's aggregation (§ 2.2) results in large writes. In fact, we observe that (§ 4) only one segment is used in Neko, which is approximately 128MiB; iPIC3D outputs smaller files, which use between 1.8 and 5.6MiB per segment.

## 9 Related Works

Caching is widely adopted to improve storage performance and characterized into two approaches. The first is write-through caching, where data is synchronously written to both the cache and primary storage. This does not help write performance, but can improve subsequent reads as they can be retrieved from the caching layer. The second is write-back caching, which is relevant to ParaLog. Rather than synchronously writing data to storage, it is first buffered in the caching layer before being asynchronously written to storage. This improves write performance as I/O operations can return quickly.

*Block-level cache.* Block-level cache is transparent to both file systems and applications. NetApp's Mercury [17] is a write-through cache where the writes are immediately flushed to the remote storage. The cache thus holds no dirty pages and serves subsequent reads. Consistent write-back caching [47] and LSVD [33], on the other hand, perform write-back caching, where a durable write returns immediately after being locally persisted, and are beneficial for write-heavy workloads.

*File system level cache.* Arion [35] proposes host-side journal for the Ceph file system [93]. Writes are buffered in DRAM, then synced to a journal backed by a block storage device. LPCC [69] extends the Lustre [80] file system to use local storage through its Hierarchical Storage Management (HSM). SymphonyFS [63], which is evaluated against ParaLog (§ 8.5), exposes a caching layer as a FUSE file system. All these approaches suffer from issues with data safety (§ 5.4) and deployability (§ 5.5). DDN's IME [77], and HadaFS [36] integrate caching at I/O forwarding nodes, target specific HPC clusters, and require dedicated hardware. This makes it challenging to deploy on small-scale on-premise clusters. Adapting them to the cloud increases cost while underutilizing already available local storage.

*Library-level cache.* Hermes [50] is a library-level approach that resides under the MPI-IO layer and redirects system calls to the local storage layer. However, it does not support crash recovery when using the write-back mode. Spectral [63, 96] is a simple write-back cache that redirects file operations to local storage and syncs data to remote after capturing the `close` system call. It does not support shared file I/O. Data Elevator [25] is a cache implemented inside the HDF5 I/O library (which also uses MPI-IO), thus only supports applications using the HDF5 file format. Similarly, ADIOS2 [30] is a streaming-oriented I/O library that can utilize local storage, but it stores data in a custom binary-packed format.

## 10 Discussion and Future Works

ParaLog shows that end-to-end completion time for data-intensive scientific applications can be improved by addressing bandwidth utilization (§ 3.1), platform environment mismatch (§ 3.2), and operational mismatch (§ 3.3). This is achieved transparently, and without compromising—even enhancing—crash consistency.

ParaLog uses a distributed, snapshot-aware log and leverages synchronization points from high-level libraries to ensure cross-node consistency. While we have demonstrated ParaLog with MPI-based scientific applications, it could support any application that uses collective communication and synchronization primitives. For example, ML frameworks such as PyTorch output and overwrite checkpoint files asynchronously. This means that an interruption will leave even the previous checkpoint file in a corrupt state [62]. With ParaLog, only consistent snapshots are checkpointed to the remote storage. As long as a synchronization point can be captured, possibly transparently (e.g, Allreduce), ParaLog can be applied. We intend to explore ParaLog in other parallel applications, such as ML frameworks, in the future.

Although ParaLog focuses on write-only workloads (§ 6.2), read-after-write (RAW) can be supported by checking the in-memory table and retrieving data from segment files. If an offset is not found, the request can be forwarded to remote storage. In this case, consistency cannot be guaranteed. For files requiring cross-process write-after-write or read-after-write, a job-level temporary file system is a better fit (§ 5.1), which can complement ParaLog's focus on write-intensive files.

## 11 Conclusion

Data-intensive applications, such as scientific workloads, are sensitive to low bandwidth and often underutilize available resources due to platform and operational mismatches. We proposed ParaLog, a distributed logging system that supports cross-node data consistency while overlapping compute-output to improve end-to-end job completion time. ParaLog accelerates three unmodified scientific applications across five platforms – including on-premise cluster, traditional HPC, and cloud HPC – covering NFS, shared PFS, and cloud-managed PFS. We show for the first time how applications can use fully-fledged MPI-IO to write data directly to S3 in an immediately shareable format. This is done on a recent production HPC system that supports S3 as a major storage subsystem, demonstrating its impact. We provided those features through three approaches: no modification in the application, no additional support from the cloud operator or system administration, and support for unmodified MPI libraries, including closed-source ones.

## Acknowledgments

This work was partly supported by the Engineering and Physical Sciences Research Council [EP/V053418/1]. We acknowledge EuroHPC Joint Undertaking for awarding us access to LUMI at CSC, Finland, and Vega at IZUM, Slovenia. The authors would like to thank Prof Adrian Jackson and EPCC for providing access and assisting us in using the NEXTGenIO system.